\documentclass[letterpaper,10pt, twoside, nosectionnumbsers]{classes/LTJournalArticle}

\def\doctitle{A simple model for entangled photon generation in resonant structures}

\def\authorOne{Nicholas J. Sorensen}
\def\authorTwo{Vitaliy Sultanov}
\def\authorThree{Maria V. Chekhova}
\def\addressOneA{
    Physics Dept. and Nexus for Quantum Technologies, University of Ottawa, 25 Templeton Street, Ottawa ON K1N 6N5, Canada
    }
\def\addressTwo{
    Max Planck Institute for the Science of Light, Staudtstr. 2, 91058 Erlangen, Germany
    }

\def\addressThree{University of Erlangen-Nürnberg, Staudtstr. 7/B2, 91058 Erlangen, Germany}

\def\emailContact{maria.chekhova@mpl.mpg.de} 

\def\abstractText{
The ability to engineer pairs of entangled photons is essential to quantum information science, and generating these states using spontaneous parametric down-conversion (SPDC) in nano- and micrometer-scale materials offers numerous advantages.
To properly engineer such sources, a reliable model describing nano- and micrometer-scale SPDC is necessary; however, such a theoretical description remains a challenge.
Here, we propose and derive a simplified model to describe SPDC in resonant structures, which considers the generation of photon pairs and the resonant enhancement of spectral bands to be separate processes, even though they actually occur simultaneously.
We compare our simplified model to both the rigorous theory of SPDC in an etalon -- a simple example of a resonant structure -- and our experiments on SPDC in etalons and find agreement for low-gain SPDC. 
By simplifying the calculations required to generate photon pairs, our model promises to make designing complex resonant structures easier, and it promises to hasten the iteration of designs across the field of quantum state engineering.
}

\usepackage{orcidlink,breqn,enumitem,caption,subcaption,ulem,titlesec,lipsum,mathtools,gensymb,float,setspace,siunitx,amsmath,multicol}

\usepackage[nameinlink]{cleveref}
\usepackage{dirtytalk}
\usepackage{upgreek}
\usepackage{dblfloatfix}

\Crefname{equation}{Eq.}{Eqs.}
\Crefname{figure}{Fig.}{Figs.}
\Crefname{table}{Tab.}{Tabs.}

\begin{document}
\renewcommand{\maketitlehookd}{
    \begin{abstract} 
    \abstractText
    \end{abstract}
}

\title{\doctitle}

\author{\normalsize\authorOne,\textsuperscript{1,2}\orcidlink{0000-0002-4666-5791} \authorTwo,\textsuperscript{2,3}\orcidlink{0000-0002-2486-4140} 
and \authorThree\textsuperscript{*,2,3}\orcidlink{0000-0002-3399-2101}}
\date{\small\textit{\textsuperscript{1}\addressOneA \\ \textsuperscript{2}\addressTwo \\ \textsuperscript{3}\addressThree \\ \textsuperscript{*}\emailContact}}

\maketitle

\section*{Introduction}

Recent years have seen significant advancement in nanoscale sources of entangled photons, which typically use spontaneous parametric down-conversion (SPDC).
These nanoscale sources include subwavelength films of strongly nonlinear material~\cite{Santiago-Cruz_OptLett_2021,Sultanov_OptLett_2022,Guo2023,Weissflog2024,Santos2024}, dielectric nanoresonators~\cite{Xu_ACSNano_2019,Marino_Optica_2019,Weissflog_LaserPhotonicsRev_2022,Weissflog2024_APR}, nanowires~\cite{Saerens_NanoLett_2023}, and resonant metasurfaces~\cite{Santiago-Cruz_NanoLett_2021,Parry_AdvPhoton_2021,Santiago-Cruz_Science_2022,Zhang_SciAdv_2022,Jia_arXiv_2024,Weissflog_Nanophotonics_2024,Noh2024}. 
The advantages of such sources are numerous; apart from compactness and integrability, some feature large spectral-~\cite{Okoth_PRL_2019,Santiago-Cruz_NanoLett_2021} and angular~\cite{Okoth2020} bandwidths and, consequently, narrow correlations in time and space. 
High degrees of continuous-variable (time/energy or position/momentum) entanglement can be achieved~\cite{Okoth_PRL_2019,Okoth2020}, as well as multiple entanglement links within the same spectrum~\cite{Santiago-Cruz_Science_2022}. 
Nanoscale sources of SPDC also enable the generation of photon pairs with tuneable polarization entanglement~\cite{Sultanov_OptLett_2022,Weissflog2024,MA_NanoLett_2023}.
Further, in extremely thin sources, photon pairs can be emitted in the opposite direction of the incident pumping radiation~\cite{Santiago-Cruz_NanoLett_2021} or they can be emitted bidirectionally, with one photon emitted forward and the other one backward~\cite{Son_Nanoscale_2023,Weissflog_Nanophotonics_2024}. 
All this makes nanoscale SPDC a revolutionary method of quantum state engineering. 

\begin{figure}[t!]
    \centering
    \includegraphics[width=0.48\textwidth]{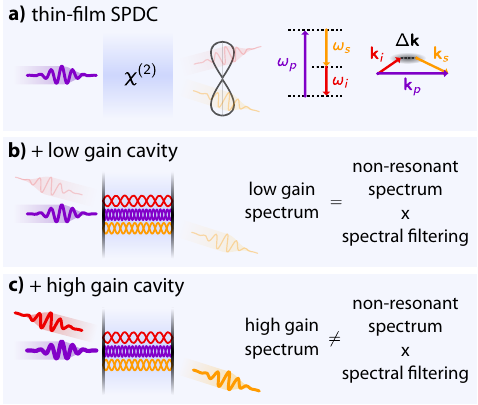}
    \caption{Decoupling photon-pair generation and spectral enhancement in nonlinear resonant structures. 
    \textbf{a)} Generation of photon pairs via SPDC in thin films is primarily dependent on energy and momentum conservation. 
    \textbf{b)} Adding a resonant structure like an etalon changes the photon pair spectrum. 
    The processes of photon-pair generation and resonant band selection can be treated sequentially in the case of low parametric gain. 
    \textbf{c)} In the case of high parametric gain, however, the two processes cannot be separated.}
    \label{fig:linearDecouplingSimple}
\end{figure}
\begin{figure*}[b!]
    \centering
    \includegraphics[width = \linewidth]{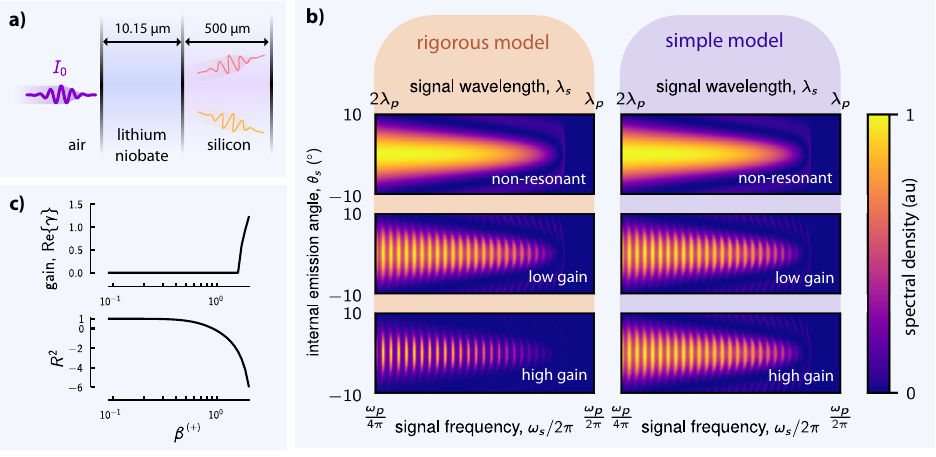}
    \caption{Comparison of rigorous and simplified models describing photon-pair generation.  
    \textbf{a)} A sketch of a 10.15$\,\upmu$m-thick lithium niobate etalon. \textbf{b)} Normalized frequency-angular spectra of photon pairs generated in the etalon sketched in (a), neglecting the boundary reflections (top) and accounting for them (bottom and middle), using a rigorous model (left) and our simplified model (right).
    In the low-gain case, the frequency-angular spectrum can be closely approximated by linearly decoupling the photon-pair generation and filtration processes.
    That is, the spectrum can be closely approximated by multiplying the non-resonant spectrum by the transmission spectrum of the etalon.
    In the high-gain regime (bottom) the approximation does not hold.
    Spectra for both the simple- and rigorous-models are calculated using frequency- and angle-dependent Fresnel coefficients and phase matching~\cite{Green2008,Small1997}.
    \textbf{c)} SPDC gain (top) and the r-squared difference between the simple and rigorous models as a function of the parametric interaction term normalized by the wavevector mismatch.
   See \Cref{app:NLModel,app:simpleModel,app:freqAngSpectrum} for details on the parameters used in the spectral calculations -- they are the same as those used in experiment.}
    \label{fig:TheoryDiagram}
\end{figure*}
Despite these advances, the theoretical description of nanoscale SPDC remains a challenge, especially in the case of metasurfaces and other nanostructured nonlinear materials. 
The standard approach to describing SPDC is based on the field quantization for plane-wave modes~\cite{Klyshko2018-fr}. 
This approach is valid in most materials because the vacuum fluctuations, which can be considered as the `seed' for SPDC, are distributed uniformly in wavevector space.
In nanostructured materials, however, `geometric' resonances enhance the vacuum field within certain frequency bands, usually found using numerical simulations. 
These numerical solutions are difficult to combine with the standard analytical approach, invalidating its use in nanostructured materials.
In recent theoretical works, a Green's function approach, developed based on the quasi-normal modes of a nanostructure, has been applied to nanoresonators~\cite{Weissflog2024_APR,Weissflog_LaserPhotonicsRev_2022}. 
However, no comparison with experiment has been made. 

Here, we propose and derive a simplified model to describe SPDC in resonant structures. 
The method is based on the fact that SPDC is a linear process under low parametric gain.
That is, the number of photons seeding the process is not amplified along the sample. 
Then, the two stages of the process -- generation of photon pairs and the selection of resonant bands -- can be considered sequentially, although, in reality, they occur simultaneously. 
To test this model, we apply it to SPDC in etalons -- thin films with considerable reflection at both faces (see \Cref{fig:linearDecouplingSimple}). 
An etalon is a one-dimensional resonant structure and can be considered the simplest case of a micro- or nano-structure. 
We compare the results to both the rigorous theory of SPDC in an etalon~\cite{Kitaeva1982,Kitaeva2004} and our experiments on SPDC in a thin nonlinear slab with highly reflective interfaces.
For low-gain SPDC, we derive the simplified model in the case of an etalon and find agreement between our simplified model and both experiment and rigorous theory.
We then sketch how our model could be applied to other resonant nonlinear structures, such as resonant metasurfaces.

\section*{Simplified model of two-photon-generation in resonant structures}

In a non-resonant nonlinear material, vacuum fluctuations are uniformly distributed in wavevector space, and the spectrum of the produced photon pairs is primarily dependent on energy and momentum conservation. 
In the case of thin nonlinear materials, the momentum mismatch does not have to be small, in accordance with the uncertainty principle. 
Therefore, non-resonant thin crystals can produce very broadband photon pairs in both frequency and direction of emission.
We calculate the intensity of photon pair emission from such a crystal, namely a 10.15$\,\upmu$m-thick layer of lithium niobate (\Cref{fig:TheoryDiagram}a), as a function of frequency and emission angle -- the frequency-angular spectrum -- and plot it in \Cref{fig:TheoryDiagram}b, top. 
The frequency-angular spectrum shown in \Cref{fig:TheoryDiagram}b demonstrates that SPDC in thin films can be very broadband. 
For details on the frequency-angular spectra calculations, see \Cref{app:freqAngSpectrum}.

If the nonlinear material is placed in a resonant structure, the structure modifies the vacuum field distribution in wavevector space.
Therefore, the spectrum of generated photon pairs depends on both the resonant structure, and energy and momentum conservation. 
Rigorously, the resonant structure modifies the photon pair spectrum in a nonlinear way, however, in the case of low parametric gain, the spectrum of photon pair emission can be expressed as 
\begin{align}
    P_{\text{res}} = P \times S. \label{eq:linearDecoupling}
\end{align}
Here, $P$ is the spectrum of photon pair emission from a non-resonant material and is material-, frequency-, and angle-dependent. 
The filtering function, $S$, depends only on the resonant properties of the structure and the collection scheme. 
The probability of detecting a photon pair depends on the coupling rates of the resonant modes to each detector and interference effects of the photon pair. 
Our model greatly simplifies the calculations needed to predict the photon pair spectrum compared to more rigorous models, though linear decoupling of the pair generation and resonant band selection does not hold in the case of high parametric gain.

Possibly the simplest case of a resonant micro- or nanostructure is the etalon. 
The space between the etalon's two interfaces supports certain modes and suppresses others, and photon pairs can be emitted forward or backward.
We calculate the probability of photon emission as a function of frequency and emission angle for a 10.15$\,\upmu$m-thick lithium niobate crystal (see the sketch in \Cref{fig:TheoryDiagram}a). 
The thickness is chosen to be large for easier comparison with the experiment; however, the same approach is valid for a nanoscale layer~\cite{Santiago-Cruz_OptLett_2021,Sultanov_OptLett_2022}.
We consider three cases: one without an etalon (the reflective interfaces are index-matched), one with an etalon with low parametric gain, and one with an etalon with high parametric gain. 
We calculate the spectra for each case using two methods and compare them in \Cref{fig:TheoryDiagram}b: the first uses a rigorous model~\cite{Kitaeva1982,Kitaeva2004} and the other one, our simplified model. 
We describe the rigorous model in \Cref{app:NLModel}, and in \Cref{app:simpleModel} we derive the form of the simplified model [\Cref{eq:linearDecoupling}] for an etalon in the low-gain regime from the rigorous one.
There is perfect agreement between our simplified model and the rigorous one in the non-resonant and low-gain cases; however, our simplified model does not hold in the high-gain case. 

In \Cref{fig:TheoryDiagram}c, we plot the r-squared ($R^2$) difference between the two models for degenerate SPDC, as well as the parametric gain as a function of $\beta^{(+)}$, where $\beta^{(+)}$ is the forward parametric interaction term (see \Cref{app:NLModel,app:simpleModel} for details).
For the simplified model to be valid, the two-photon generation process must occur in the low-gain regime, requiring that $\left[\beta^{(+)}\right]^2\ll1$.
In that regime, we find perfect agreement between our simplified model and the more rigorous one, as demonstrated by the $R^2$ value plotted in \Cref{fig:TheoryDiagram}c. 
In the case of high gain, where $\beta^{(+)}$ is large, the spectrum is modulated strongly, and our simplified model is no longer commensurate with the rigorous one.
In the high-gain regime, $\beta^{(+)}$ can also start to dominate the phase mismatch, causing $\text{Re}\{\gamma^{(+)}\}$ to become large (see \Cref{fig:TheoryDiagram}c, top).

\section*{Experimental generation of photon pairs in an etalon}

\begin{figure*}[t!]
    \centering
    \includegraphics[width=\linewidth]{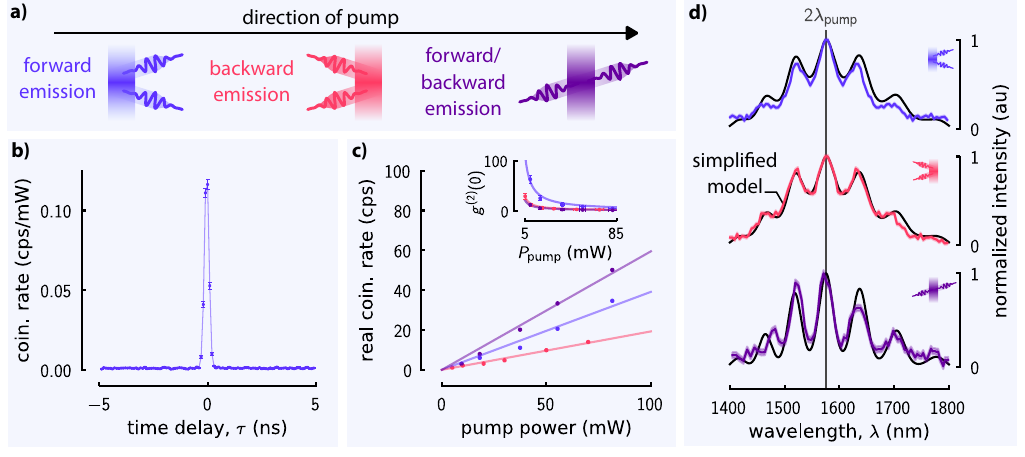}
    \caption{Detection of photon pairs generated in a lithium niobate etalon. 
    \textbf{a)} Different measurement schemes, shown by different colors. 
    \textbf{b)} Distribution of the time delay between the arrivals of the two photons for the case of forward emission. 
    \textbf{c)} Real coincidence rate as a function of the $\SI{788}{nm}$-wavelength pump power. 
    The inset gives the second-order correlation function $g^{(2)}(0)$ as a function of pump power. The $g^{(2)}(0)$ values are low due to the high levels of photoluminescence from the silicon layer. 
    \textbf{d)} The spectrum of the detected photon pairs in each detection scheme, where the Poissonian measurement uncertainty is given by the lighter colored line for each data set. 
    See \Cref{fig:timeDelaySpectra} for the raw time-delay spectra and \Cref{fig:rawSpectra} for a comparison of the measured coincidence rates to those predicted by the simplified model.
    Each spectrum is plotted against the one predicted by our simplified model, which considers the wavelength-dispersive collection and detection efficiencies of our experimental setup using a fitted gaussian envelope function. 
    The vertical gray line denotes the degenerate signal and idler wavelengths.}
    \label{fig:figSiMeasurements}
\end{figure*}
To verify our approach, we measure the spectrum of photon pairs produced in a nonlinear etalon comprised of a $10.15\,\upmu$m-thick $x$-cut lithium niobate (LN) crystal on a silicon substrate (see \Cref{app:experimentalSetupDesc} for details).
There is a large difference in refractive index between the three materials (air, LN, and silicon), which creates two reflective interfaces and can behave resonantly at both the pump and photon pair wavelengths.
Using this structure, we measure the spectra of photon pairs emitted in the forward and backward directions and compare them to our simplified model. 

We pump SPDC with a 788\,nm continuous-wave (CW) laser with up to 70\,mW power, focused to a spot diameter of 5\,$\upmu$m inside the LN crystal. 
The pump is incident on the LN side of the sample to minimize the photoluminescence inside the silicon. 
We use a Hanbury-Brown-Twiss setup to detect photon pairs emitted in both the forward (parallel to the pump) and backward (antiparallel to the pump) directions (see \Cref{fig:figSiMeasurements}a and \Cref{app:experimentalSetupDesc} for details).
In both directions, we collect the photons emitted from the LN, send them through fiber beamsplitters, and detect them using two superconducting nanowire single-photon detectors. 
We analyze the delays between the two detection times, which results in a pronounced peak of simultaneous detections (real coincidences, see \Cref{fig:figSiMeasurements}b). The background is due to accidental coincidences involving unpaired photons, mainly from the photoluminescence.  
The rate of real coincidences is proportional to the pump power (\Cref{fig:figSiMeasurements}c), and the ratio of the total and accidental coincidence detection rates [given by the second-order correlation function at zero delay $g^{(2)}(0)$] is inversely proportional to the pump power (\Cref{fig:figSiMeasurements}c, inset).
The data shown in Figs.\,\ref{fig:figSiMeasurements}b and \ref{fig:figSiMeasurements}c evidence the generation of entangled photon pairs.

Next, we use two-photon fiber spectroscopy to measure the spectrum of the photon pairs~\cite{Valencia2002}. 
A three kilometer-long SMF-28 optical fiber, which behaves as a wavelength-dependent time delay between different photons, is added between the crystal and the detectors.
Longer wavelength photons travel with a different group velocity than shorter wavelength photons, and we can convert the difference in photon arrival time into wavelength.
Using this method, we measure the spectra of photon pairs emitted forward and backward and plot them in \Cref{fig:figSiMeasurements}d. 
We compare the measured spectra to our simplified model and find good agreement. 
In this case, we include the wavelength-dependent detector efficiency in our modeling.
These experiments demonstrate the usefulness of our model, which eases the complexity of spectral calculations of photon-pair generation at low parametric gain.

To extend the simplified model to other micro- and nanostructures, such as metasurfaces, we posit that it is sufficient to understand how the structure modifies resonant fields and how they out-couple to radiated fields. 
This knowledge would give the form of $S$ in \Cref{eq:linearDecoupling}, given that the structures operate in the low-gain regime and constitute a single nonlinear source.
Further experimental studies could help to refine our model and verify its accuracy for other resonant structures.

\section*{Conclusion}
In this paper, we have developed and validated a simplified model to describe spontaneous parametric down-conversion (SPDC) in resonant nanostructures. 
Using the fact that, under low parametric gain, SPDC is a linear process, our model decouples photon-pair generation from resonant mode selection, providing a straightforward and effective method to predict photon pair emission spectra.
Perfect agreement was found when comparing our simplified model to the rigorous theory of SPDC in an etalon, which we derive it from, and the simplified model closely predicts the emission direction and spectrum of photon pairs produced by an etalon. 

By reducing computational complexity, the simplified model introduced here will enable iteration and optimization of nanophotonic structures like resonant metasurfaces, nanowires, and dielectric nanoresonators. 
The model, however, only works in the case of low parametric gain, as nonlinear interactions in high-gain regimes lead to discrepancies. 
Further, the general model has only been delineated for the case of an etalon; future experimental studies on a wider variety of resonant nanostructures could refine the model, extending its validity to a broader range of conditions, including deep subwavelength confinement. 
Nevertheless, this work promises to streamline the design of complex resonant structures used to generate photon pairs and other quantum states, facilitating advances in quantum state engineering.

\appendix
\vspace{1cm}
\noindent\large{\bfseries{Appendices}}
\normalsize
\section{Nonlinear model describing two-photon-generation in an etalon}
\label{app:NLModel}
\begin{figure}[b!]
    \centering
    \includegraphics{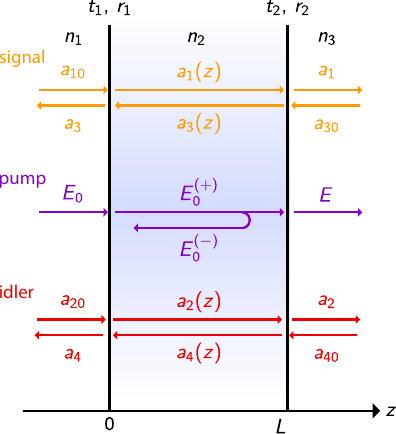}
    \caption{Geometry of the waves involved in SPDC in an etalon. The $\mathbf{z}$-direction defines the incident direction of the pump wave.
    Though the signal and idler modes are depicted parallel to the $\mathbf{z}$-axis, we consider angled modes inside the resonator by calculating the angle-dependent Fresnel coefficients and phase mismatch.}
    \label{fig:ScatteringMatrixDefinitions}
\end{figure}
Let us first consider a rigorous approach for the description of the etalon effect on photon-pair generation in a thin layer of nonlinear material~\cite{Kitaeva1982, Kitaeva2004}.
Specifically, we use the scattering matrix approach to calculate the efficiency of photon-pair generation in a nonlinear layer of thickness L, sandwiched between other materials.
The generally dispersive refractive indices of each non-absorbing layer are given by $n_i = n_i(\omega)$ for $i\in\{1,2,3\}$, and the transmission (reflection) coefficients for the first and second interfaces are $t_1$ ($r_1$) and $t_2$ ($r_2$), respectively.
These coefficients are frequency- and angle-dispersive, and can be calculated using the Fresnel formulas~\cite{hecht2016}.
We consider three modes propagating inside and outside the etalon: the pump (purple), the signal (orange), and the idler (red) (\Cref{fig:ScatteringMatrixDefinitions}), and each frequency mode inside the layer has two corresponding spatial modes propagating forward or backward. 
The forward- ($E_0^{(+)}$) and backward-propagating ($E_0^{(-)}$) pump modes inside the layer are excited by a single pump mode ($E_0$), incident from the left side of the structure, whereas
the modes of the down-converted photons inside the layer, combined into the array $\mathbf{A}(z) = \{\hat{a}_1(z), \hat{a}_2^{\dagger}(z), \hat{a}_3(z), \hat{a}_4^{\dagger}(z)\}$, are pair-wise coupled through the nonlinear interaction inside the layer due to the second-order nonlinearity.  

We assume the pump field is incident along the $\mathbf{z}$-axis; however, we place no such restrictions on the photon pairs.
To account for angled modes propagating in the resonator, we calculate the angle-dependent Fresnel coefficients and phase matching.
In the case of birefringent media, we also assume that the parametric interaction involves only signal and idler modes with definite polarization such that birefringent effects vanish~\cite{Kitaeva2004}. 
Further, to avoid the effects of birefringence, we assume that any angled signal and idler modes exist in a single plane.
With these assumptions, the polarizations of linearly polarized incident and reflected waves remain invariant.
Further, to simplify calculations, the idler modes are already Hermitian-conjugate.

To determine the emission spectra of the nonlinear layer, we have to consider all possible combinations of the corresponding spatial modes.
There are eight possible combinations: for each pump mode inside the layer, there are four down-converted spatial modes: both photons are emitted forward (ff), both are emitted backward (bb), one is emitted forward and the other backward (fb), and vice-versa (bf). 
The spatial modes of each down-converted photon, $\hat{a}_i(z)$, are coupled to the corresponding input mode, $\hat{a}_{i0}$, and output mode, $\hat{a}_{\text{i}}$, by the boundary conditions at each interface.
Then, we solve for the array of field operators $\mathbf{A}_{\text{out}} = \{a_1, a_2^{\dagger}, a_3, a_4^{\dagger}\}$ of the emitted fields, given the array of the incident field operators $\mathbf{A}_{\text{in}} = \{a_{10}, a_{20}^{\dagger}, a_{30}, a_{40}^{\dagger}\}$.
To write the boundary conditions, we first define the array of field operators before propagating through the layer, $\mathbf{A}' = \{\hat{a}_1(0), \hat{a}_2^{\dagger}(0), \hat{a}_3(L), \hat{a}_4^{\dagger}(L)\}$
, and after propagating through it, $\mathbf{A}'' = \{a_1(L), a_2^{\dagger}(L), a_3(0), a_4^{\dagger}(0)\}$.
Hence, the boundary conditions of the field amplitudes can be written as
\begin{align}
    \mathbf{A}'&=\hat{\rho} \mathbf{A}'' + \hat{\tau}_1 \mathbf{A}_{\text{in}},  \\ \mathbf{A}_{\text{out}}&=\hat{\tau}_2 \mathbf{A}''-\hat{\rho}^{\dagger} \mathbf{A}_{\text{in}}, \label{eq:BC}
\end{align}
where the transmission and reflection matrices are respectively given by 
\begin{align}
    \hat{\tau}_1& = 
    \begin{bmatrix}
    t_{1}^{(s)} & 0 & 0 & 0\\
    0 & t_{1}^{(i)*} & 0 & 0\\
    0 & 0 & t_{2}^{(s)} & 0\\
    0 & 0 & 0 & t_{2}^{(i)*}
    \end{bmatrix}, \\
    \hat{\tau}_2 &= 
    \begin{bmatrix}
    t_{2}^{(s)} & 0 & 0 & 0\\
    0 & t_{2}^{(i)*} & 0 & 0\\
    0 & 0 & t_{1}^{(s)} & 0\\
    0 & 0 & 0 & t_{1}^{(i)*}
    \end{bmatrix},\\
    \hat{\rho}& = 
    \begin{bmatrix}
    0 & 0 & r_{1}^{(s)}e^{i\phi_s} & 0\\
    0 & 0 & 0 & r_{1}^{(i)*}e^{-i\phi_i}\\
    r_{2}^{(s)}e^{i\phi_s} & 0 & 0 & 0\\
    0 & r_{2}^{(i)*}e^{-i\phi_i} & 0 & 0
    \end{bmatrix}, 
\end{align}
and $t_{m}^{(\mu)}$ and $r_{m}^{(\mu)}$ are the transmission and reflection coefficients of the $m$-th interface ($m\in\{1,2\}$) at frequency $\omega_\mu$ ($\mu\in\{s,i\}$).
The phase accumulated by each mode as it propagates through the slab ($\phi_\mu = Lk_\mu^\parallel$) is captured in these matrices. 
Here, $k_\mu^\parallel$ is the magnitude of the parallel component of the wavevector of the $\mu$-th mode inside the etalon. 
We assume that the slab is thick enough so that the phase matching still does not allow direct generation of counter-propagating photons.
Finally, the pump field is modified by the etalon:
\begin{align}
    E_{0}^{(+)}&=\frac{t^{\text{(p)}}_{1}}{1-r^{\text{(p)}}_1 r^{\text{(p)}}_{2} \exp \left(2 i \phi_{\text{p}}\right)} E_{0},  \\
    E_{0}^{(-)}&=\frac{t^{\text{(p)}}_{1} r^{\text{(p)}}_{2} \exp \left(i \phi_{\text{p}}\right)}{{1-r^{\text{(p)}}_1}r^{\text{(p)}}_{2} \exp \left(2 i \phi_{\text{p}}\right)} \\E_{0}&=r^{\text{(p)}}_{2} \exp \left(i \phi_{\text{p}}\right) E_{0}^{(+)}.
\end{align}

Now, the interaction between each field inside the etalon can be represented by an interaction matrix $\hat{w}$, such that 
\begin{align}
    &\mathbf{A}'' = \hat{w}\mathbf{A}',  
    \\
    &\hat{w} = 
    \begin{bmatrix}
    w_{11}^{(+)} & w_{12}^{(+)} & 0 & 0\\
    w_{21}^{(+)} & w_{22}^{(+)} & 0 & 0\\
    0 & 0 & w_{11}^{(-)} & w_{12}^{(-)}\\
    0 & 0 & w_{21}^{(-)} & w_{22}^{(-)}
    \end{bmatrix}\label{eq:interactionMatrix},
\end{align} 
where 
\begin{align}
& w^{(\pm)}_{11}=e^{-i\Delta/2}\left(\cosh \gamma^{(\pm)}+\frac{i\Delta}{2}\frac{ \sinh \gamma^{(\pm)}}{\gamma}\right), \label{eq:w11}\\
& w^{(\pm)}_{22}=e^{i\Delta/2}\left(\cosh \gamma^{(\pm)}-\frac{i\Delta}{2}\frac{ \sinh \gamma^{(\pm)}}{\gamma}\right), \label{eq:w22}\\
& w^{(\pm)}_{12}=-i \beta^{(\pm)} \frac{\sinh \gamma^{(\pm)}}{\gamma^{(\pm)}},  \label{eq:w12}\\
&w^{(\pm)}_{21}=i \beta^{(\pm)} \frac{\sinh \gamma^{(\pm)}}{\gamma^{(\pm)}}. \label{eq:w21}
\end{align}
Here, $\Delta \coloneqq L\Delta k^{\parallel}$ is the phase mismatch along the pump axis inside the etalon ($\Delta k^{\parallel} = k_{\text{p}}^\parallel - k_{\text{s}}^\parallel - k_{\text{i}}^\parallel $), 
\begin{align}
    \beta^{(\pm)} \coloneqq \frac{2 \pi \omega_1 \omega_2 \chi^{(2)} L E^{(\pm)}_0}{c^2 \sqrt{k_{\text{s}}^\parallel k_{\text{i}}^\parallel}} \label{eq:ParametricInteractionTerm}
\end{align} 
is the parametric interaction term, $c$ is the speed of light in vacuum, and 
\begin{align}
    \gamma^{(\pm)} = \left(\left[\beta^{(\pm)}\right]^2  - \frac{\Delta^2}{4}\right)^{1/2} 
\end{align}
is the gain term. 
Here, $(+)$ denotes the gain corresponding to the forward traveling pump field amplitude, and $(-)$ denotes the opposite.

Details on these equations can be found in \cite{Kitaeva2004}. 
Using \Cref{eq:BC,eq:interactionMatrix}, we can define the scattering matrix $\hat{U}$, given by $\mathbf{A}_{\text{out}} = \hat{U}\mathbf{A}_{\text{in}}$ such that
\begin{align}
    \hat{U} = \hat{\tau_2} \hat{w} (\hat{I}-\hat{\rho}\hat{w})^{-1} \hat{\tau_1} - \hat{\rho}^{\dagger}. \label{eq:scatteringMatrix}
\end{align}
Now, \Cref{eq:scatteringMatrix} can be used to calculate the emitted field amplitudes. 
More specifically, we find that the probability that two photons will be emitted forward is $ \tilde{P}_{\text{res}}^{\text{ff}}  = \left\langle \text{vac}\right| a_2^\dagger a_1^\dagger a_1 a_2\left| \text{vac}\right\rangle$,  
\begin{align}
\begin{split}
    \tilde{P}_{\text{res}}^{\text{ff}} \propto \langle \text{vac} | 
    &(U_{21} \hat{a}_{10} + U_{22}\hat{a}_{20}^\dagger+ U_{23} \hat{a}_{30}  + U_{24}\hat{a}_{40}^\dagger)\\ 
    \times&(U_{11}^* \hat{a}_{10}^\dagger + U_{12}^* \hat{a}_{20}+ U_{13}^* \hat{a}_{30}^\dagger  + U_{14}^* \hat{a}_{40})\\
    \times&(U_{11} \hat{a}_{10} + U_{12}\hat{a}_{20}^\dagger+ U_{13} \hat{a}_{30}  + U_{14}\hat{a}_{40}^\dagger)\\
    \times&(U_{21}^* \hat{a}_{10}^\dagger + U_{22}^* \hat{a}_{20}+ U_{23}^* \hat{a}_{30}^\dagger  + U_{24}^* \hat{a}_{40}) | \text{vac} \rangle.
\end{split} 
\end{align}
Simplifying this expression, we find that 
\begin{align}
\begin{split}
    \tilde{P}_{\text{res}}^{\text{ff}} \propto &|U_{21}|^2 \left(|U_{11}|^2 + |U_{12}|^2 + |U_{14}|^2 \right)   \\
    &+ |U_{23}|^2 \left(|U_{13}|^2 + |U_{12}|^2 + |U_{14}|^2 \right) \\
    &+ 2\operatorname{Re}\{U_{11}U_{23}U_{21}^* U_{13}^*\}.
\end{split} \label{eq:St} 
\end{align}
Similarly, we find expressions for the probability that two photons will be emitted backward, $\tilde{P}_{\text{bb}}$, and that one photon will be emitted forward and the other reflected backward, $\tilde{P}_{\text{fb}}$, and vice-versa, $\tilde{P}_{\text{bf}}$:
\begin{align}
\begin{split}
    \tilde{P}_{\text{res}}^{\text{bb}} \propto &|U_{41}|^2 \left(|U_{31}|^2 + |U_{32}|^2 + |U_{34}|^2 \right)   \\
    &+ |U_{43}|^2 \left(|U_{33}|^2 + |U_{32}|^2 + |U_{34}|^2 \right) \\
    &+ 2\operatorname{Re}\{U_{31}U_{43}U_{41}^* U_{33}^* \};
\end{split} \label{eq:Sr} \\
\begin{split}
    \tilde{P}_{\text{res}}^{\text{fb}} \propto &|U_{41}|^2 \left(|U_{11}|^2 + |U_{12}|^2 + |U_{14}|^2 \right)   \\
    &+ |U_{43}|^2\left(|U_{12}|^2 + |U_{13}|^2 + |U_{14}|^2 \right) \\
    &+ 2\operatorname{Re}\{U_{41}U_{13}U_{11}^* U_{43}^* \}.
\end{split} \label{eq:Str} \\
\begin{split}
    \tilde{P}_{\text{res}}^{\text{bf}} \propto &|U_{21}|^2 \left(|U_{31}|^2 + |U_{32}|^2 + |U_{34}|^2 \right)   \\
    &+ |U_{23}|^2 \left(|U_{33}|^2 + |U_{32}|^2 + |U_{34}|^2 \right)\\
    &+ 2\operatorname{Re}\{U_{31}U_{23}U_{21}^* U_{33}^* \}.
\end{split} \label{eq:Srt} 
\end{align}
The accented probability, $\tilde{P}_{\text{res}}$, is used to differentiate the probability predicted by the rigorous model from the one predicted by the simplified model, $P_{\text{res}}$, introduced in the main text and discussed in the following section. The rigorous spectra shown in \Cref{fig:TheoryDiagram}c are calculated using \Cref{eq:St}. 
To simplify those calculations, we assume that the idler emission angle minimizes the transverse wavevector mismatch (see \Cref{app:freqAngSpectrum}).
Further, we calculate the angle-dispersive transmission and reflection coefficients calculated using the Fresnel formulas and the experimentally measured refractive indices of silicon and lithium niobate \cite{Small1997,Green2008}. 

\section{Derivation of the simple model describing two-photon generation in an etalon}
\label{app:simpleModel}
While the rigorous quantum formalism expounded above offers a rigorous description of SPDC with the etalon effect, it can be computationally intensive and complex. 
However, under specific conditions, it is possible to instead use a simplified semi-classical model. 
Specifically, in the low-gain regime, photon-pair generation can be linearly decoupled from the filtration effects of a resonator. 
The two processes -- generation and filtration -- can be treated multiplicatively. 
The generation of the photon pairs can be modeled identically to that in a non-resonant thin film, while the spectral filtering effect of the etalon can be included as a modulation of the photon pair spectrum.
We derive this simplified model from the rigorous one presented in \Cref{app:NLModel}.
The spectral filtering effect, $S$, (from \Cref{eq:linearDecoupling} in the main text) is described here, whereas the non-resonant probability of generating a photon pair, $P$, is described in \Cref{app:freqAngSpectrum}.

The simplified model describing two-photon generation in an etalon requires that the process occurs in the low-gain regime. 
This requirement places a condition on the parametric interaction term [\Cref{eq:ParametricInteractionTerm}], i.e. $|\beta^{(\pm)}|^2\ll1$, and this primarily limits the range of resonant pump intensities for which the model holds.
This assumption allows us to simplify the gain term, $\gamma^{(\pm)}$. 
To do so, we consider two detuning limits and derive the form of the scattering matrix, which is identical for both limits. 
In the case of large detuning ($\Delta^2/4 \gg |\beta^{(\pm)}|^2$), we can rewrite the gain term as $\gamma^{(\pm)} \approx i\Delta/2$.
This assumption then allows us to simplify the interaction matrix terms introduced in \Cref{eq:w11,eq:w22,eq:w12,eq:w21}:
\begin{align}
& w^{(\pm)}_{11}\approx e^{-i\Delta/2}\left(\cosh \frac{i\Delta}{2}+\sinh \frac{i\Delta}{2}\right)=1, \label{eq:w11s}\\
& w^{(\pm)}_{22}\approx e^{i\Delta/2}\left(\cosh \frac{i\Delta}{2}-\sinh \frac{i\Delta}{2}\right)=1, \label{eq:w22s}\\
& w^{(\pm)}_{12}\approx - \beta^{(\pm)} \frac{\sinh \frac{i\Delta}{2}}{\frac{\Delta}{2}}=- \beta^{(\pm)}\text{sinc}\frac{\Delta}{2},  \label{eq:w12s}\\
&w^{(\pm)}_{21}\approx \beta^{(\pm)} \frac{\sinh \frac{i\Delta}{2}}{\frac{\Delta}{2}}= \beta^{(\pm)}\text{sinc}\frac{\Delta}{2}. \label{eq:w21s}
\end{align}
It is easy to show that these equations also hold in the case that $\Delta/2\ll 1$. 
Therefore, we can write the simplified interaction matrix as 
\setlength{\arraycolsep}{-2pt}
\begin{align}
\hat{w} \approx  
    \begin{bmatrix}
    1 & -\beta^{(+)}\text{sinc}\frac{\Delta}{2} & 0 & 0\\
    \beta^{(+)}\text{sinc}\frac{\Delta}{2} & 1 & 0 & 0\\
    0 & 0 & 1 & - \beta^{(-)}\text{sinc}\frac{\Delta}{2}\\
    0 & 0 & \beta^{(-)}\text{sinc}\frac{\Delta}{2} & 1
    \end{bmatrix}\label{sssss}.
\end{align} 
Now, we can calculate a simpler version of the scattering matrix using \Cref{eq:scatteringMatrix}, and the spectral densities associated with different emission directions can be calculated using \Cref{eq:St,eq:Sr,eq:Str,eq:Srt}. 
The resultant large analytical equations can be algebraically reduced to simple, digestible expressions, assuming that $\beta^{(\pm)}\ll1$. 
When fully expanded, the analytical expressions contain terms scaled by various powers of $\left[\beta^{(\pm)}\right]^2$ -- terms scaled to higher order than $\left[\beta^{(\pm)}\right]^2$ are ignored. The simplified model is therefore accurate to $\mathcal{O}\left(\left[\beta^{(\pm)}\right]^4\right)$.

In the case of two forward-emitted photons, the simple model predicts that the spectral density of two forward-emitted photons can be written as
\begin{align}
P^{\text{ff}}_{\text{res}}&\propto \underbrace{\text{sinc}^2\left(\frac{\Delta}{2}\right)}_{P} \underbrace{\Bigr| \beta^{(+)} a_1^{(+)} a_2^{(+)}+\beta^{(-)} a_1^{(-)} a_2^{(-)} \Bigr| ^2}_{S} ,\label{eq:Pffs}
\end{align}
where we have expressed the equation in the form suggested by \Cref{eq:linearDecoupling} -- $P$ is the spectrum of photon pair emission from a non-resonant material and $S$ is the filtering function of the resonant material. 
The form of $P$ is discussed further in \Cref{app:freqAngSpectrum}.
The parametric interaction terms for the forward- $(+)$ and backward-traveling $(-)$ pump fields are the same as given in \Cref{eq:ParametricInteractionTerm}.
The etalon-enhanced field amplitudes of the down-converted photons, $a_i$, are also the same as before (\Cref{fig:ScatteringMatrixDefinitions}), where the additional superscripts denote the direction of the pump inside the layer:
\begin{align}
    a_{1,2}^{(+)} &= \frac{t_2^{(\text{s,i})}}{1 - r_1^{(\text{s,i})}r_2^{(\text{s,i})}e^{i2\phi_{\text{s,i}}}}, \label{eq:a12p}\\ 
    a_{1,2}^{(-)} &= \frac{r_1^{(\text{s,i})}t_2^{(\text{s,i})}e^{i\phi_{\text{s,i}}}}{1 - r_1^{(\text{s,i})}r_2^{(\text{s,i})}e^{i2\phi_{\text{s,i}}}}, \label{eq:a12m}\\    
    a_{3,4}^{(+)} &= \frac{r_2^{(\text{s,i})}t_1^{(\text{s,i})}e^{i\phi_{\text{s,i}}}}{1 - r_1^{(\text{s,i})}r_2^{(\text{s,i})}e^{i2\phi_{\text{s,i}}}}, \label{eq:a34p}\\ 
    a_{3,4}^{(-)} &= \frac{t_1^{(\text{s,i})}}{1 - r_1^{(\text{s,i})}r_2^{(\text{s,i})}e^{i2\phi_{\text{s,i}}}}.    \label{eq:a34m}
\end{align}

Each field amplitude is etalon-enhanced and multiplied by reflection and transmission coefficients that correspond to the pump field direction and emission direction.
In a similar fashion, we determine the spectral densities of the other collection schemes to be
\begin{align}
P^{\text{bb}}_{\text{res}}&\propto\text{sinc}^2\left(\frac{\Delta}{2}\right) \Bigr| \beta^{(+)} a_3^{(+)} a_4^{(+)}+\beta^{(-)} a_3^{(-)} a_4^{(-)} \Bigr| ^2,\label{eq:Pbbs}\\
P^{\text{fb}}_{\text{res}}&\propto \text{sinc}^2\left(\frac{\Delta}{2}\right) \Bigr| \beta^{(+)} a_1^{(+)} a_4^{(+)}+\beta^{(-)} a_1^{(-)} a_4^{(-)} \Bigr| ^2,\label{eq:Pfbs}\\
P^{\text{bf}}_{\text{res}}&\propto \text{sinc}^2\left(\frac{\Delta}{2}\right) \Bigr| \beta^{(+)} a_3^{(+)} a_2^{(+)}+\beta^{(-)} a_3^{(-)} a_2^{(-)} \Bigr| ^2.\label{eq:Pbfs}
\end{align}

The simplified model constituted by \Cref{eq:Pffs,eq:Pbbs,eq:Pfbs,eq:Pbfs} is of the same form as \Cref{eq:linearDecoupling} and validates the linear, sequential treatment of photon pair generation and the selection of resonant bands in etalons.
The same equations [\Cref{eq:Pffs,eq:Pbbs,eq:Pfbs,eq:Pbfs}] are used to predict the experimental spectra plotted in \Cref{fig:figSiMeasurements} in the main text. For those plots, we assume that all photons are emitted along the $\mathbf{z}$-axis, as integrating across all emission angles returned virtually identical results.

We must also note that the simplified model derived here and constituted by \Cref{eq:linearDecoupling} is only valid for a single nonlinear source (i.e. the photon pair must co-propagate with the pump field). 
For instance, in the case of extremely thin two-photon generation sources in which the permitted momentum mismatch is sufficient to allow for counter-propagating photon pairs, the co-propagating- and counter-propagating-pairs are generally described by different non-resonant photon emission spectra.
For this counter-propagating photon pairs to be non-negligible, however, the thickness of the films must be deeply subwavelength and would require exotic materials and structures~\cite{Santiago-Cruz_OptLett_2021}. 
To describe these kinds of systems, a more general model is required. 

\section{Calculating the frequency-angular spectrum of SPDC in thin films}
\label{app:freqAngSpectrum}
Next, we calculate the spectrum of SPDC in non-resonant thin films.
In a nonlinear material of length $L$ and second-order susceptibility $\chi^{(2)}$, the probability of producing a photon pair is given by 
\begin{align}
P \propto \left| F_{\text{pm}}\left(\Delta k^{\parallel}\right) F_{\mathrm{p}}\left(\Delta k^{\perp}\right)\right|^2, \label{eq:FASThinFilm}
\end{align}
where $F_{\text{pm}}$ is the phase-matching function, and $F_{\text{p}}$ is the pump function. 
This is the same $P$ as in \Cref{eq:linearDecoupling}.
The wavevector mismatch components are given by
\begin{align}
&\Delta k^{\parallel}:=k^{\parallel}_{\text{p}}-k^{\parallel}_{\text{s}}-k^{\parallel}_{\text{i}} ; \\
&\Delta k^{\perp}:=k^{\perp}_{\text{p}}-k^{\perp}_{\text{s}}-k^{\perp}_{\text{i}} .
\end{align}
We define the longitudinal direction to be along the pump ($z$) direction. The transverse direction is in the plane of the crystal ($x$). The phase matching and pump functions are given by 
\begin{align}
    F_{\mathrm{pm}}\left(\Delta k^{\|}\right)&=\operatorname{sinc}\left(\frac{\Delta k^{\|} L}{2}\right)\exp\left({i\frac{\Delta k^{\|} L}{2}}\right), \\
    F_{\mathrm{p}}\left(\Delta k^{\perp}\right)&=e^{-\left(\Delta k^{\perp} w/2\right)^2}, 
\end{align}
respectively. 
Here, $w$ is the $1/e^2$ pump waist diameter.
In \Cref{fig:TheoryDiagram}, we employ the same waist diameter as used in experiment, $w=\SI{5}{\upmu m}$, and we calculate the transmission and reflection coefficients using the Fresnel formulas and the experimentally measured refractive indices of silicon and lithium niobate~\cite{Small1997,Green2008}.

In \Cref{fig:TheoryDiagram}b, we plot the frequency-angular spectra for different structures as a function of only the signal frequency and internal emission angle. 
For all such calculations we assume that the frequency of the idler photon is chosen to satisfy energy conservation, and for simplicity we assume that its angle is chosen to minimize the transverse wavevector mismatch, $\Delta k^{\perp}$.
For a strongly focused pump, a more accurate calculation might integrate the spectral emission over all possible idler emission angles $\theta_i$ for a given $\theta_s$ to produce the frequency-angular spectrum.

\section{Experimental setup used to generate and detect photon pairs}
\label{app:experimentalSetupDesc}
\begin{figure*}[t!]
    \centering
    \includegraphics[width = \textwidth]{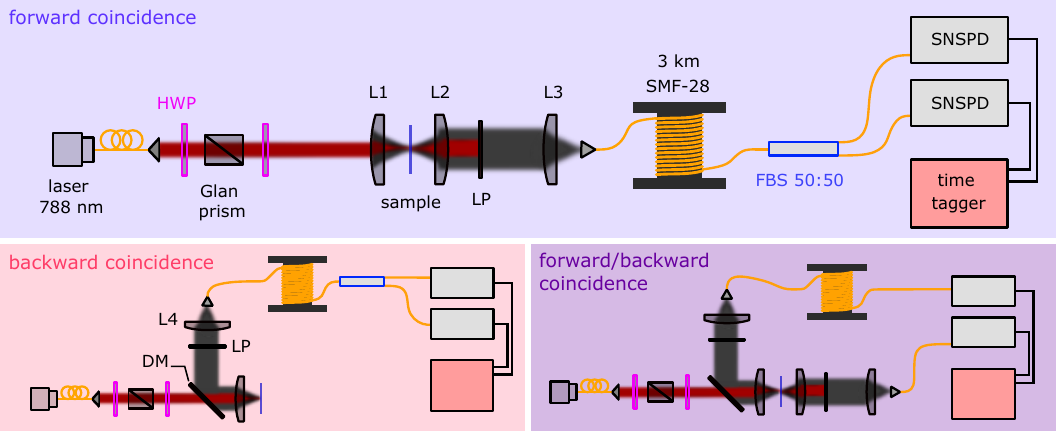}
    \caption{Experimental setup used to generate and detect the photon pairs: DM -- dichroic mirror, short pass; FBS -- fiber beamsplitter; HWP -- half-wave plate; L1 -- lens, 15\,mm focal length (actually a 15\,mm parabolic mirror); L2 -- lens, 60\,mm focal length; L3 -- lens, 30\,mm focal length; L4 -- lens, 11\,mm focal length; LP -- long-pass filter; SNSPD -- superconducting nanowire single photon detector.
    }
    \label{fig:fiberSpectroscopySetup}
\end{figure*}
\begin{figure*}[b!]
    \centering
    \includegraphics[width=\linewidth]{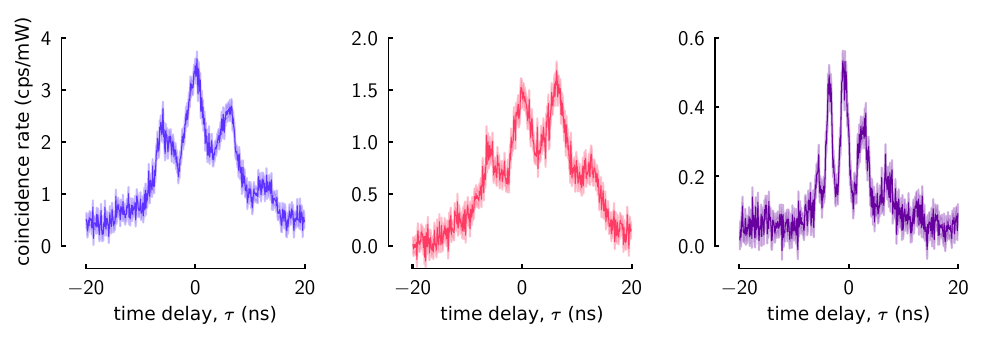}
    \caption{The raw, time-delayed coincidence measurements in the three measurement schemes: forward (left), backward (middle), and forward/backward (right). }
    \label{fig:timeDelaySpectra}
\end{figure*}
We use the modified Hanbury-Brown-Twiss experimental setups shown in \Cref{fig:fiberSpectroscopySetup}, where photon pairs are produced by focusing the pump into the sample. 
The photons are collimated, filtered, split on a fiber beamsplitter, and sent to two superconducting nanowire single photon detectors (SNSPDs) connected to a time tagger. 
For all correlation measurements, we use a continuous wave (CW) laser with center wavelength of $\lambda =$~788\,nm and a 2\,nm full-width at half-maximum (FWHM) bandwidth, which produces up to 100\,mW of power (model LP785-SF100 from Thorlabs). 
The laser has a $1/e^2$ diameter in air of 3.0\,mm and is focused to a spot with a diameter of 5$\,\upmu$m using a parabolic mirror with a 15\,mm focal length.
The pump is polarized along the $z$ crystal axis.

We use two SNSPDs (NW1fC900-1550 detectors from Photon Spot) to detect the photon pairs. 
The detectors are single-mode fiber (SMF) coupled and have detection efficiencies above 80\,\% for 900$\,\sim\,$1500\,nm light (the actual detection efficiency is lower and has a lower bandwidth because of dispersive loss in the SMF). 
The detectors have dark counts at 100$\,\sim\,$300\,Hz and jitters $<\,$100\,ps at FWHM. 
To filter out the pump and photoluminescence, we use several 1000\,nm longpass (LP) filters and a 1400\,nm LP filter.

\begin{figure}[t!]
    \centering
    \includegraphics[width = 0.48\textwidth]{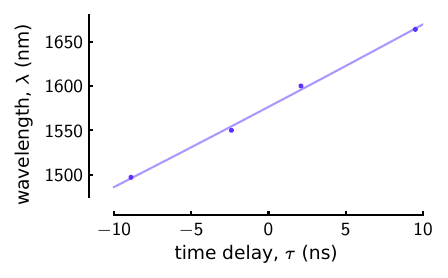}
    \caption{Calculated regression (solid line) from a series of filtered measurements (scatter points).}
    \label{fig:fiberDispersionRegression}
\end{figure}

\begin{figure*}[ht!]
    \centering
    \includegraphics[width=\linewidth]{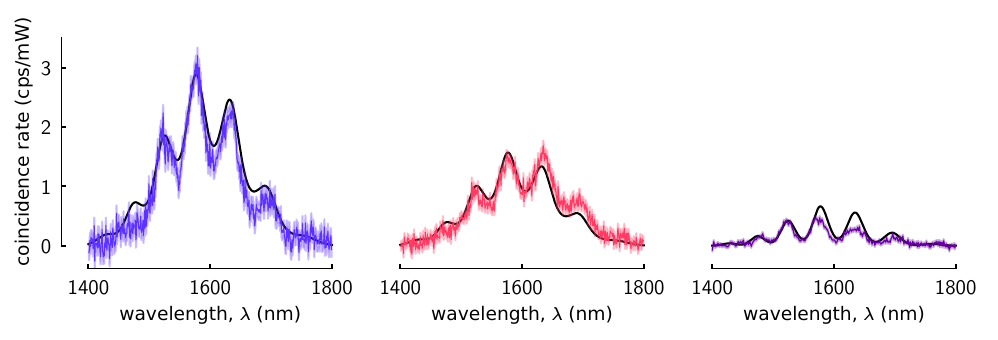}
    \caption{The raw spectral coincidence measurements showing the relative coincidence rates for each of the three measurement schemes: forward (left), backward (middle), and forward/backward (right). 
    We also plot the relative coincidence rates predicted by the simplified model (black), normalized to the maximum coincidence rate in the forward scheme and estimated collection efficiencies. 
    See the text for more information. } 
    \label{fig:rawSpectra}
\end{figure*}

Three kilometers of SMF are added between the fiber couplers and one detector (in the forward/backward detection scheme) or both detectors (in the forward and backward detection schemes) to temporally delay the photon arrival by dispersion, producing the data shown in \Cref{fig:timeDelaySpectra}. 
This SMF is removed for the measurements presented in \Cref{fig:figSiMeasurements}b and \Cref{fig:figSiMeasurements}c. 
The effective refractive index of the normally dispersive SMF varies approximately linearly as a function of wavelength. 
The measured temporal correlations are then linearly related to the wavelength of the radiation.
The dispersion of SMF is not widely reported, so we calibrate the system using a broadband source ($x$-cut lithium niobate) and a series of bandpass (BP) and LP filters. 
This produces a series of measurements with well-defined edges from which time delay-wavelength correlations are made. 
From these measurements, a regression is fit (\Cref{fig:fiberDispersionRegression}). 
Then, the rate of photon pair detection can be plotted as a function of wavelength, as shown in \Cref{fig:figSiMeasurements}d and \Cref{fig:rawSpectra}.
    
Featuring the same data presented in \Cref{fig:figSiMeasurements}d, \Cref{fig:rawSpectra} emphasizes that the measured coincidence rates and spectra match those predicted by the simplified model in all three detection schemes. 
The data presented in \Cref{fig:rawSpectra} has not been averaged, as was done in \Cref{fig:figSiMeasurements}d. 
Sample placement and optics were not altered between the three spectral measurements and should therefore have the same collection rates, assuming there was negligible drift between the three experiments. 
However, despite attempts to maximize photon free-space coupling to fiber, there were differences in photon collection efficiencies between the forward- and backward-emitted photons. 
We scale the simplified model predictions plotted in \Cref{fig:rawSpectra} by these collection efficiencies, which we estimate to be $\eta_f/\eta_b=0.4$. 
The forward- and backward-collection efficiencies, $\eta_f$ and $\eta_b$, respectively, were estimated by inspection of the raw spectral coincidence measurements.

\label{sec:refs}

\subsection*{Funding}
\vspace{-0.2cm}
This work was funded by Deutsche Forschungsgemeinschaft (DFG, German Research Foundation), Project 311185701, and we acknowledge support from the Canada Research Chairs Program, the Natural Sciences and Engineering Research Council of Canada, as well as the Canada First Research Excellence Fund.

\subsection*{Acknowledgments}
\vspace{-0.2cm}
We sincerely thank Jeff S. Lundeen for his discussions and his help in procuring some of the samples used in this work. 
We also thank Changjin Son for useful conversations.
Collaboration between groups at the Max Planck Institute and the University of Ottawa was made possible through the Ottawa-Erlangen Max-Planck Centre.

\subsection*{Disclosures}
\vspace{-0.2cm}
The authors declare no conflicts of interest.

\subsection*{Data availability}
\vspace{-0.2cm}
Experimental data underlying the results presented in this paper are not publicly available at this time but may be obtained from the authors upon reasonable request.
Code used to generate the manuscript's non-experimental plots is available at Ref.~\cite{Sorensen2025}.

\bibliography{bibliographies/main}
\end{document}